\def\year{2022}\relax
\documentclass[letterpaper]{article} 
\usepackage{aaai22}  
\usepackage{times}  
\usepackage{helvet}  
\usepackage{courier}  
\usepackage[hyphens]{url}  
\usepackage{graphicx} 
\usepackage{natbib}  
\usepackage{caption} 
\DeclareCaptionStyle{ruled}{labelfont=normalfont,labelsep=colon,strut=off} 
\usepackage{algorithm}
\usepackage{algorithmic}

\usepackage{array}
\usepackage{makecell}
\usepackage{xcolor}

\usepackage{hyperref}
\usepackage{pdfpages}

\usepackage{newfloat}
\usepackage{listings}
\floatstyle{ruled}
\newfloat{listing}{tb}{lst}{}
\floatname{listing}{Listing}
\title{Experiences of Censorship on TikTok Across Marginalised Identities}
\author{
    Eddie L. Ungless, \textsuperscript{\rm 1}\\\
    Nina Markl, \textsuperscript{\rm 2}\\
    Bj\"{o}rn Ross \textsuperscript{\rm 1}\\
}
\affiliations{
    \textsuperscript{\rm 1}University of Edinburgh\\
    \textsuperscript{\rm 2}University of Essex\\

    Correspondence: b.ross@ed.ac.uk
}

\begin{document}

\maketitle

\begin{abstract}
TikTok has seen exponential growth as a platform, fuelled by the success of its proprietary recommender algorithm which serves tailored content to every user - though not without controversy. Users complain of their content being unfairly suppressed by ``the algorithm'', particularly users with marginalised identities such as LGBTQ+ users. Together with content removal, this suppression acts to censor what is shared on the platform. Journalists have revealed biases in automatic censorship, as well as human moderation. We investigate experiences of censorship on TikTok, across users marginalised by their gender, LGBTQ+ identity, disability or ethnicity. We survey 627 UK-based TikTok users and find that marginalised users often feel they are subject to censorship for content that does not violate community guidelines. We highlight many avenues for future research into censorship on TikTok, with a focus on users' folk theories, which greatly shape their experiences of the platform.

\end{abstract}

\section{Introduction}
In 2024, TikTok is one of the largest social media networks in the world, having amassed over 1 billion users faster than any app ever before \citep{Harwell_2022}. It serves machine-learning curated, user-generated video content; a format that has proven so successful that other social platforms are following suit \citep{Harwell_2022, Frier_2022}. For some younger users, TikTok has taken the place of Google as a source of news and local information \citep{Harwell_2022}. Despite its popularity, users' experiences are far from universally positive: complaints about censorship on the platform, including the removal or suppression of content that does not seem to violate TikTok’s community guidelines, are common (e.g. \citet{Brown_2021,Karizat_Delmonaco_Eslami_Andalibi_2021, Are_2023}). In particular, hashtags and terms used by minority groups appear to be affected \citep{Ryan_Fritz_Impiombato_2020}. 

(Perceived) censorship on TikTok of content by marginalised creators has received significant media coverage \citep{Brown_2021,Lorenz_2022,kelion_2019, ohlheiser_2021,kover_reuter_2019,Biddle_2020}, including a Netzpolitik article which shared leaked documents which revealed that content featuring marginalised individuals such as those with disabilities is deliberately suppressed by moderators of the platform \citep{kover_reuter_2019}, a practice which seems to have continued despite push-back \citep{Biddle_2020}. The topic has also been explored through surveys \citep{Haimson_Delmonaco_Nie_Wegner_2021} and interviews with users \citep{Karizat_Delmonaco_Eslami_Andalibi_2021, Simpson_Semaan_2021, Are_2023, 10.1145/3613904.3642148}, which surfaced users' beliefs about how moderation is conducted. Whilst these may not align with the platforms' actual moderation system, they provide us with useful insight into the motivations behind users' behaviours when they interact with this system \citep{Karizat_Delmonaco_Eslami_Andalibi_2021}. 

Understanding marginalised users' experiences of censorship on TikTok is vital because it is a widely used platform that serves to connect marginalised creators with their community whilst at the same time policing what they share, as \citet{Simpson_Semaan_2021} highlight for LGBTQ+ users. This allows TikTok to play an unprecedented role in the creation of online and offline identities. Biased censorship also alienates already marginalised individuals, reflecting offline power imbalances \citep{Are_2023}. Surveying users allows us to better understand how they interpret this censorship. 

We use censorship as an umbrella term covering both removal and suppression of content, by the recommender algorithm and the moderation system(s) (two systems\footnote{\url{https://www.tiktok.com/transparency/en/content-moderation/} and \url{https://newsroom.tiktok.com/en-us/how-tiktok-recommends-videos-for-you}} which are often conflated by TikTok users and referred to as ``the algorithm'') and by human moderators, though our focus is on algorithmic censorship. Automated video removal now accounts for the largest proportion of removed videos: over 100M every quarter.\footnote{\url{https://www.tiktok.com/transparency/en/community-guidelines-enforcement-2023-2/}} Suppression refers to when a creator's content has its ``reach'' limited, for example by no longer being served to viewers, though it is still on the creator's profile. This has also been referred to as ``shadow banning'' when it happens systematically \citep{Are_2022, Are_2023}. We ask respondents for their experiences of content removal and suppression, their beliefs with regards to algorithmic censorship and how this might reflect or differ from community guidelines, and the roles that human moderators might play. By gathering detailed demographic data, we can explore how experiences of censorship differ across identities. 

In the following, we present the necessary background to understanding our findings, including an overview of TikTok and a review of the literature into perceived social media censorship. We present our survey methodology, then present descriptive findings and statistical analyses and discuss these results. Our exploratory work highlights many avenues for future research. The explosive increase in TikTok usage has left researchers and policy makers to catch up in our understanding of user experiences, a vital step in knowing how to protect users from harms related to fairness, censorship and knowledge manipulation. Our quantitative analyses complement existing qualitative work on the topic of marginalised users' experiences on TikTok. Beyond existing quantitative work, we consider both removal and (suspected) suppression of content, the latter being particularly relevant to a platform where the user experience is so heavily driven by algorithmically curated content. This work makes an important contribution to our understanding of (perceived) fairness of censorship on the TikTok, on a scale not achieved by existing work.

\section{Background}
\subsection{Introduction to TikTok}
TikTok is a platform for sharing short-form video content; it describes itself as ``the leading destination'' for this content.\footnote{\url{https://www.tiktok.com/about?lang=en}} 
Like many social media platforms, TikTok employs an algorithm to curate content shown to users, and uses automated moderation, including algorithms, to filter out inappropriate content.\footnote{\label{com}\url{https://www.tiktok.com/transparency/en/community-guidelines-enforcement-2022-2/}} TikTok has published community guidelines which outline what content is inappropriate for the platform, including sexual content and content promoting violence.\footnote{\url{https://www.tiktok.com/community-guidelines?lang=en}}

Content moderation is primarily presented as a way to keep users safe, and as TikTok claim ``to foster a fun and inclusive environment''.\textsuperscript{\ref{com}} However, as \citet{Cobbe_2021} describes, automated moderation allows ``unprecedented... control'' over users' public and private content. \citet{Zeng_Kaye_2022} have referred to the algorithmic suppression of content on TikTok as ``visibility moderation''. The recommender ``For You'' algorithm can be used to enforce moderation decisions, for example by not serving content flagged as ``shocking... to a general audience''.\footnote{\label{how}\url{https://newsroom.tiktok.com/en-us/how-tiktok-recommends-videos-for-you}} The recommender system may also learn to suppress content that would otherwise not be subject to moderation, for example by not serving content from LGBTQ+ creators because it is similar to content that has previously received a lot of negative interaction. Thus the recommender ``For You'' algorithm and moderation systems\footnote{\url{https://www.tiktok.com/transparency/en/content-moderation/}}\textsuperscript{,\ref{how}} can both be said to conduct algorithmic censorship, understanding censorship to include both removal and suppression of content. Indeed,  TikTok users often seem to conflate these into a single entity known as ``the algorithm'', responsible for controlling the reach of content as well as removal: for example, users talk of content being ``taken down'' by the algorithm \citep{Karizat_Delmonaco_Eslami_Andalibi_2021}. In this paper, we discuss users' attitudes towards this algorithmic censorship.

\subsection{Biased moderation on social media}
Whilst ostensibly intended to protect communities, moderation on social media can also enact harm. Studies into moderation on social media suggest that marginalised users face additional censorship. \citet{Haimson_Delmonaco_Nie_Wegner_2021} surveyed users of several social media platforms and noticed a trend whereby Black and trans users reported having content removed that related to their marginalised experiences, even though the content followed site policies. ``Conservative'' users also reported high levels of removal, but this was typically related to violations of the platform's policies on hate speech or misinformation. Whilst all three groups may feel the platform shows a bias against them, the content removed from conservative users was often in violation of platform policies -- instances of true positives -- whereas Black and trans respondents reported high levels of false positives. Therefore the moderation systems on these platforms can be said to be biased against Black and trans creators.

Further, moderation systems can be exploited by malicious agents to enact harm on marginalised creators, for example by reporting content to get a creator silenced or banned \citep{Zeng_Kaye_2022, Are_2023}. \citet{Are_2023} highlights how already marginalised creators such as sex workers and LGBTQ+ individuals can face targeted campaigns of reporting as a form of harassment.

\subsection{Harms of (biased) algorithmic censorship}
Considering briefly the negative impact automated censorship on social media can have even if it were operating in an unbiased manner, \citet{Cobbe_2021} argues that automated censorship allows privately owned social platforms' commercial priorities (the desire to feature content that appeals to the mainstream) to be ``inserted'' into the private and public conversations of platform users, undermining ``open and inclusive discussion''. This may prevent users from expressing themselves in an authentic manner, as they are prompted to align what they express with the commercial goals of the platform. 

A censorship algorithm that is (perceived to be) biased can cause further harm at an individual and societal level. Discriminatory censorship can lead to both representational and allocation harms against a community, following the distinction made by \citet{barocas2017problem}. A scenario where a representational harm might occur is when the censorship algorithm removes content featuring self-described fat creators \citep{Clark_Lee_Jingree_2021}, reinforcing the fatphobic belief that only thin bodies should be seen. Algorithms regulate ``what becomes visible and what remains out of sight'' \citep{Velkova_Kaun_2021}. An allocational harm might occur if a user's content being censored harms their income, as would be the case for the many influencers and small businesses who rely on social media. Further, the suppression and removal of posts relating to ``Black Lives Matter'', reported by users of TikTok \citep{ghaffary_2021}, could be argued to deny users' the opportunity to contribute to protests on the platform.

\citet{Ehsan_Singh_Metcalf_Riedl_2022} write in reference to an unfair grading algorithm that ``algorithms can leave imprints on how people make sense of algorithmic operations and interpret their lived experiences with the algorithm, carrying deep psychological impact on their mental well-being.''. Having content unfairly censored by the app, or observing this happening, could lead to feelings of alienation and lack of agency, feelings which may remain after dis-use \citep{Ehsan_Singh_Metcalf_Riedl_2022}. 

\subsection{Folk theories of censorship}
\citet{Karizat_Delmonaco_Eslami_Andalibi_2021} found that users of TikTok believed that ``the algorithm'' suppressed content based on the creators' social identity, understood as referring to one's membership in a certain social group \citep{Burke_Stets_2009}. More specifically, users felt that content was suppressed based on race and ethnicity, physical appearance including body size, disability and class status, LGBTQ+ identity and political/ social justice group affiliations. Those belonging to marginalised groups had their content suppressed, whereas others -- those with ``algorithmic privilege'' \citep{Karizat_Delmonaco_Eslami_Andalibi_2021} -- benefitted from having their content favoured by the platform. This finding accords with work by \citet{Simpson_Semaan_2021} who found that LGBTQ+ users of TikTok felt that TikTok unfairly censored content posted by them and fellow LGBTQ+ creators, all whilst pigeon-holing them as belonging to (normative) queer identities in terms of the content they were served. The algorithm constructed a profile for these users based on an approximation of their queer identities, which determined what they could see, and what they could post.

The belief that the platform censors content based on social identity is referred to as The Identity Strainer Theory, an example of one of several folk theories \citet{Karizat_Delmonaco_Eslami_Andalibi_2021} argue users hold about content curation and moderation. Such folk theories shape the way users behave on social media platforms such as TikTok \citep{West_2018,Karizat_Delmonaco_Eslami_Andalibi_2021, Are_2023}, as well as their interactions with other technologies such as smart devices \citep{Frick_Wilms_Brachten_Hetjens_Stieglitz_Ross_2021} or their response to algorithmic grading \citep{Ehsan_Singh_Metcalf_Riedl_2022}. Folk theories can develop as a result of ``everyday algorithm auditing'', whereby users detect problematic behaviour through their every day interactions with a system \citep{Shen_DeVos_Eslami_Holstein_2021}, as is the case when users of TikTok deliberately interact with certain content to determine what the algorithm prioritises \citep{Karizat_Delmonaco_Eslami_Andalibi_2021, Simpson_Semaan_2021}; trial different text to determine what the app censors \citep{Brown_2021}; or observe which videos receive limited views \cite{Lyu_Cai_Callis_Cotter_Carroll_2024}.


\section{Methodology}
\subsection{Respondents}
All respondents were recruited using Prolific.com, which pseudo-anonymises all data. All respondents were UK-based. Prolific allows researchers to screen for use of TikTok. Recruitment occurred from 2023-2024. We first conducted a screening study to find respondents who had ever posted on TikTok, as we were interested in direct experiences of censorship. 2,350 respondents completed our initial prescreen, for a fee of £0.10 (equivalent to £10.59/hr). Of these, 777 completed our main study, for a fee of £1.70 (equivalent to £7.37/hr). 45 additional respondents were rejected in line with Prolific's policies (i.e. failure of multiple attention checks, answers contrary to pre-screening data). Next, we targeted LGBTQ+ respondents to ``up-sample'' queer users of TikTok. 101 LGBTQ+ respondents completed our prescreen; 50 of these completed the main study. One additional respondent was rejected. Of the 827 respondents, we used data from 627 respondents who passed all three attention checks to ensure quality responses.\footnote{Respondents were asked to select ``yes'' if they were paying attention. Two of the rating questions included an instruction to ``select somewhat disagree'' or ``select somewhat agree'', respectively} This is a high rejection rate, but as our attention checks were very simple we felt failure of even one would indicate poor quality data.

\subsection{Procedure and Measurements}
We conducted a survey for an exploratory analysis of marginalised users' experiences of censorship compared to non-marginalised users. Ethics approval was obtained from the University of Edinburgh Informatics Research Ethics Process, rt \#6862. This study was conducted online using Qualtrics.com which has excellent security protocols, and data was analysed on a password-protected computer. 

\subsubsection{Use of TikTok} After giving informed consent, respondents were asked about their use of TikTok. If respondents indicated that they had stopped using TikTok, we asked their motivations: options were based on \citet{Grandhi_Plotnick_Hiltz_2019,Vaterlaus_Winter_2021,Lu_Lu_Liu_2020, Zhou_Yang_Jin_2018}. Most relevant to the present study, we asked whether ``too much moderation / censorship'' was a motivation for leaving. Respondents could also give other motivations. 

Respondents were asked how often they used TikTok to view, and separately to post content, from ``I have never [posted/viewed] content'' to ``$>$3 times a day'' (options based on \citet{Lu_Lu_Liu_2020}). Respondents who indicated that they had never posted content were removed from our final data. We asked about the types of content respondents viewed and, separately, posted on TikTok, with categories taken from \citet{Vaterlaus_Winter_2021}. We confirmed respondents consumed and posted content in English and told them we were only interested in their use of TikTok for English language content.

\subsubsection{Experience of Censorship} We asked about respondents' experiences of censorship. We stated that by censorship ``we mean both when content is removed and when content is suppressed.'' We gave the example of a video getting very few views as something that might indicate suppression, as this is reflective of what TikTok users state they use as ``evidence'' that they have been suppressed or ``shadow banned''  \citep{Lyu_Cai_Callis_Cotter_Carroll_2024}. Removed content is no longer be visible on the platform. Use of the term ``censorship'' may have influenced our respondents - see \textit{Limitations}.

We provided a list of 13 topics (henceforth ``controversial topics''\footnote{We use this term as users have previously reported having such content removed and thus it can be thought of as ``controversial''}) and the option to supply ``other''. The list of topics was derived from \citet{Haimson_Delmonaco_Nie_Wegner_2021}'s paper on social media censorship, where users were invited to share what kind of content they felt was censored across different social media platforms. Our list is as follows: Political content; Content some may find offensive or inappropriate; Sex related content for a non-erotic purpose i.e. that is intended to educate; Sex related content for an erotic purpose; Covid-related content; Content insulting or criticizing dominant group (e.g., men, white people); Content relating to a social justice movement, for example feminism or anti-racism; Content relating to minority identity experience i.e. queer content, content about Black experiences; Hate speech; Curse words; Self-referential use of slur i.e. d*ke by a lesbian; Content about violence that is not intended to shock or disgust i.e. reporting on a violent crime; Content about violence that is intended to shock or disgust. By providing a pre-defined list we reduced the cognitive effort for respondents providing answers, and ensured we had consistent data across respondents. 

Respondents were asked about which of these topic types they had posted, and how frequently. Separately (see \textit{Limitations}) respondents could indicate whether they had had content removed, and how often these different kinds of content were removed on a 5-point scale of ``never'' to ``always''. We repeated this process for content suppression. We then asked questions relevant to a topic not explored in this paper. 

\subsubsection{Beliefs about Algorithmic Censorship}

All respondents were then asked how strongly they agreed that the TikTok moderation algorithm censors at least some posts about the 13 controversial topics, on a scale of 1-5 (``strongly disagree'' - ``strongly agree''). 

\subsubsection{Beliefs about Community Guidelines}
Respondents were asked how strongly they agreed that the TikTok community guidelines do not allow posts about the 13 controversial topics, on a scale of 1-5. 

We then explained that some users feel that content that seems to be in line with the community guidelines is censored. We asked respondents how strongly they agreed on a scale of 1-5 with the following statements about why content is censored that does not go against community guidelines: ``Other user(s) have reported the content'', ``The algorithm has not learned to follow the guidelines (it is not a good algorithm)'',  ``TikTok has unpublished guidelines that are used to train the algorithm which are stricter'', ``TikTok has unpublished guidelines that are given to human moderators which are stricter'', ``The algorithm has misunderstood the content / made a one-time mistake'', ``Human moderators have their own opinions about what should be allowed on the platform''. \citet{West_2018} found human intervention and moderators having their own biases were two primary reasons people gave for content moderation.  

\subsubsection{Demographic Information} Respondents were asked demographic questions (always with the option not to answer - ``prefer not to say''). Respondents were asked their age, their gender identity (male, female, non-binary, other [text entry]), their sexuality (straight, gay, bisexual, asexual, other [text entry]),\footnote{These were categories of sexuality. For the full list of options, see Supplementary Material} and their ethnic group (topline categories taken from the UK 2021 census \citep{census}). Respondents were asked about their trans status and disability status (yes, no). Respondents were also asked about political beliefs, socio-economic status and religion, which we do not analyse in this paper. Respondents were asked if they belonged to any other marginalised groups, and asked to give their first language(s). Participants were asked additional questions not relevant to this analysis, then debriefed. 
\section{Results}
\subsubsection{Demographic Information}
\begin{table}[]
\setlength\extrarowheight{2pt} 
    \centering
    \begin{tabular}{lrr}
        \textbf{Identity} & \textbf{Marginalised: n, \%} & \textbf{Non-marg.: n, \%} \\
        \hline
        Gender & \Gape[2pt]{\makecell[r]{Female: $377$, $60\%$ \\ Nonbinary: $24$, $6\%$ \\ Other: $1$, $0\%$}} & Male: $224$, $36\%$ \\
        \hline
        Trans & Yes: $20$, $3\%$ & No: $599$, $96\%$ \\
        \hline
        Sexuality & \Gape[2pt]{\makecell[r]{Asexual: $8$, $1\%$ \\ Bisexual: $99$, $16\%$ \\ Gay: $44$, $7\%$ \\ Other: $4$, $1\%$}} & Straight: $461$, $74\%$ \\
        \hline
        Disability & Yes: $73$, $12\%$ & No: $542$, $87\%$ \\
        \hline
        Ethnicity & \Gape[2pt]{\makecell[r]{Asian: $45$, $7\%$ \\ Black: $45$, $7\%$ \\  Multiple: $21$, $3\%$ \\ Other: $3$, $0\%$}} & White: $511$, $82\%$ \\
        \hline
    \end{tabular}
    \caption{Table showing count and percentage of respondents of marginalised and non-marg(inalised) identities. Percentages will not sum to $100$ as ``Prefer not to say'' excluded.
    }
    \label{tab:my_label}
\end{table}
Although demographic questions were asked at the end of the survey we present this data first as they contextualise the following results. The modal age was 23, the mean was 30. The reported ages skewed heavily towards the 18-30 range. This is reflective of TikTok UK user trends \citep{tiktokagegendergender_2021}. Other demographic information is given in Table \ref{tab:my_label}. There being more females is typical of TikTok's UK user base \citep{tiktokagegendergender_2021}.  The rate of LGB+ identities ($24.7\%$) is much higher than recent England and Wales census data would anticipate ($6.9\%$ \citet{census}),\footnote{Latest Scottish census data unavailable at time of writing} even before we performed up sampling of LGBT+ identities ($21\%$). This may partly be due undercount in the census \cite{KevinGuyan_2022}), but it also seems likely this is reflective of TikTok's position as a prominent social network for LGBTQ+ youth, noted elsewhere \citep{Ohlheiser_2020}. The proportion of disabled respondents is lower than census data for England and Wales \citep{census}: the audio-visual nature of the platform will have influenced this, but it is worth noting that TikTok has been shown to be biased against disabled users which may have discouraged use by disabled creators \cite{Lyu_Cai_Callis_Cotter_Carroll_2024,kover_reuter_2019,Biddle_2020}. We were unable to establish if ethnicity data is reflective of the typical UK user base of TikTok, but it is relatively reflective of the UK population, per England and Wales census data \cite{census}. We asked respondents if they belonged to any other marginalised identities and responses included ``working class'', ``neurodivergent'' and ``ex-sex worker''.

\subsection{All Respondents}
We first present the results across all respondents, to paint a picture of typical use. To ensure our results across all demographics are indicative of the typical TikTok population, we use raking to weight the sexuality groups\footnote{The ANES raking variable selection algorithm determined changes to the sample based on gender and trans status were negligible} per their prevalence in the \textit{original} recruitment drive,\footnote{For lack of more detailed TikTok user demographic information being available, we take our original sample distribution to be reasonably accurate} rather than after up-sampling of LGBTQ+ respondents. Typically, percentages are given as is appropriate for weighted data, but where it improves clarity we also report raw $n$ counts. We then analyse results across each demographic axis in turn, with a focus on how the experiences of marginalised users differ from non-marginalised users. 

\subsubsection{Use of TikTok} 
The vast majority ($90.0\%$) reported having started using TikTok over a year ago.  The majority ($55.1\%$) of respondents used TikTok to view content over 3 times a day. The modal total time spent on the app was between 1-2 hours. This suggests our data is representative of ``loyal'' users of the app. The majority ($74.1\%$) of respondents posted 0-3 times a month. This suggests most users of the app are relatively passive, consuming content but infrequently posting. 

A very small number ($3\%$) of respondents reported being ex-users of the platform. Of these, the most common reason for quitting was because content was no longer entertaining (reported by $55\%$ of ex-users). ``Too much moderation/ censorship'' was reported by a single ex-user (raw count), suggesting this is not a primary motivating factor for leaving. 
 



\subsubsection{Experience of Censorship}\label{expcensor}

The least common content type that respondents posted was ``hate speech'' ($3.15\%$). The most common controversial content types were ``Curse words'' ($35.1\%$ of respondents), followed by ``Political content'' ($19.3\%$). 

A relatively small percentage of respondents reported having had content removed ($12.8\%$) or suppressed ($14.1\%$). Around half of respondents who reported content suppression believed they had had content removed, and vice versa. Of those who report content removal, ``Other'' and ``Content some may find offensive'' account for the majority of reports of removal (reported by $38.1\%$ and $33.5\%$ respectively; respondents could select multiple content types). Of the ``Other'' content which had been removed, written answers included unwarranted concerns over minor safety (most common reason given), copyright infringement, drugs or alcohol, and content incorrectly identified as sexually inappropriate. That ``Content some may find offensive or inappropriate'' was the second most common suggests that respondents were aware (at least retroactively) that the content could offend. The most likely content type to ``Always'' be removed was violence.

For content suppression, the most common types of content respondents believed had been suppressed were ``Other'' ($26.0\%$ of those who reported suppression) then ``Curse words'' ($25.6\%$). Respondents who selected ``Other'' frequently reported being unsure why their content was being suppressed ($\sim1/3$). Compared to content removal, reports of suppression were more evenly distributed across topics. This suggests that the controversial content types are all considered possible reasons for suppression, even if removal is relatively uncommon for some e.g. Covid-related content. The most consistently suppressed content was ``Sex related content for a non-erotic purpose''.

The high number of fill in text answers suggests that the controversial topics we had identified through existing research explain only part of TikTok users' experiences of censorship on the platform (see \textit{Limitations})

\subsubsection{Beliefs about Algorithmic Censorship}\label{sec:beliefac}

Respondents most strongly agreed that ``Content about violence that is intended to shock or disgust'' is censored, selecting on average ``Somewhat agree'' ($3.88$).  Respondents were least likely to agree that ``Covid-related content'' is censored, selecting ``Neither agree nor disagree'' on average ($2.89$). 

We invited respondents to add anything else they believe about the TikTok moderation algorithm, and some themes emerged. Two respondents (raw counts) suggested the community guidelines were not being followed by the algorithmic censorship -- for example, ``They don’t follow their own community guidelines because they ban innocent people all the time but keep people who break it for e.g. I always see nudity on my fyp''. Many reported that content was removed without reason or that sometimes the ``wrong'' content was removed ($n=12$). 
Some respondents ($n=5$) shared that they felt the censorship was biased against marginalised groups, for example saying 
``Minorities are disproportionately affected... A black person can duet a racist person explaining that they are wrong, and the black person’s video will be removed... not the racist’s''. 
Several respondents ($n=3$) felt that moderation was politically motivated, for example because of its failure to remove propaganda. 


\subsubsection{Beliefs about Community Guidelines}\label{sec:beliefcg}
There were several topics that users on average agreed were censored by the algorithm, but which they thought did not go against TikTok community guidelines (or were unsure). This was true for non-erotic sex related content, curse words and political content. Of particular relevance to the focus of this paper, this was true for both content relating to minority identity experiences and content relating to a social justice movement. Respondents somewhat agreed ($4.05$) that when content was removed despite not violating community guidelines, this was due to other users reporting the content. This was the reason that respondents considered to be the most likely. 

\subsection{Experience of Censorship by Demographic Group}
\begin{table}[] \centering 
\begin{tabular}{wl{2.4cm}wl{2.4cm}wl{2.4cm}} 
\\\hline 
 & Removal & Suppression \\ 
\hline \\[-1.8ex] 
 Male & $-1.458$ (0.768) & $-$0.469 (0.593) \\ 
 Straight & $-0.883^{*}$ (0.345) & $-$0.608$^{*}$ (0.352) \\ 
 White & \makecell[c]{--} & $-$0.658$^{*}$ (0.281) \\ 
 Not Disabled & $-0.850^{*}$ (0.331)& $-$1.004$^{*}$ (0.326) \\ 
 Male + Straight & $+2.223^{*}$ (0.826) & $+1.320^{*}$ (0.664) \\ 
 Constant & $-0.766^{*}$ (0.296) & $-$0.302 (0.389) \\ 
\hline \\
\end{tabular} 
\caption{Table showing coefficients and (standard errors) in two logistic regressions predicting content removal, and content suppression. $^{*}p<.05$.}\label{tab:reg}
\end{table}

We now consider the impact of identity on experiences of censorship. We conducted logistic regressions to determine if identity predicts content removal and suppression. We exclude those who answered ``Prefer not to say'' to any demographic question ($n=25$, $4.0\%$ of data). We created dummy variables across gender, trans status, sexuality, disability and ethnicity (Marginalised by gender = 0, not marginalised by gender = 1, etc). We include binary interaction effects e.g. male + straight. We use a stepwise algorithm to determine the final model. We found disability status, gender and sexuality all impact content removal, $R^2=.05, p<.001$: straight people and people without disabilities are less likely to experience removal, but straight men specifically are more likely to experience removal (see Table \ref{tab:reg}). We found disability status, ethnicity, gender and sexuality all predict content suppression, $R^2=.06, p<.001$: white people, straight people and those without disabilities are less likely to experience suppression, but straight men are more likely. 

Identity clearly impacts experiences of censorship, and it is not always the case that those of marginalised identities experience more censorship (i.e. straight men are more likely to report censorship). We now look at demographic attributes in turn to better understand experiences of censorship across identities. 

Given the exploratory nature of this work, we did not formulate specific hypotheses, but we do perform some post-hoc testing to suggest whether differences are substantive. We always exclude ``Other'' due to small $n$ and lack of homogeneity. We primarily conduct Fisher's Exact Tests to establish the significance of the difference in rates of posting controversial content or having it removed or suppressed, between groups. We report 2-sided $p$-values\footnote{All $p$-values are from Fisher's Exact Tests unless specified} and make suggestions as to whether the differences we report are likely to be meaningful. Where relevant we include other statistical analyses, noting that typically only large effect sizes can be detected. Tests were selected which were suitable to the lack of group size parity. This exploratory paper can indicate fruitful lines of future enquiry to reify our descriptive findings.

\subsubsection{By Gender}

\begin{figure}[t]
\includegraphics[scale=0.5]{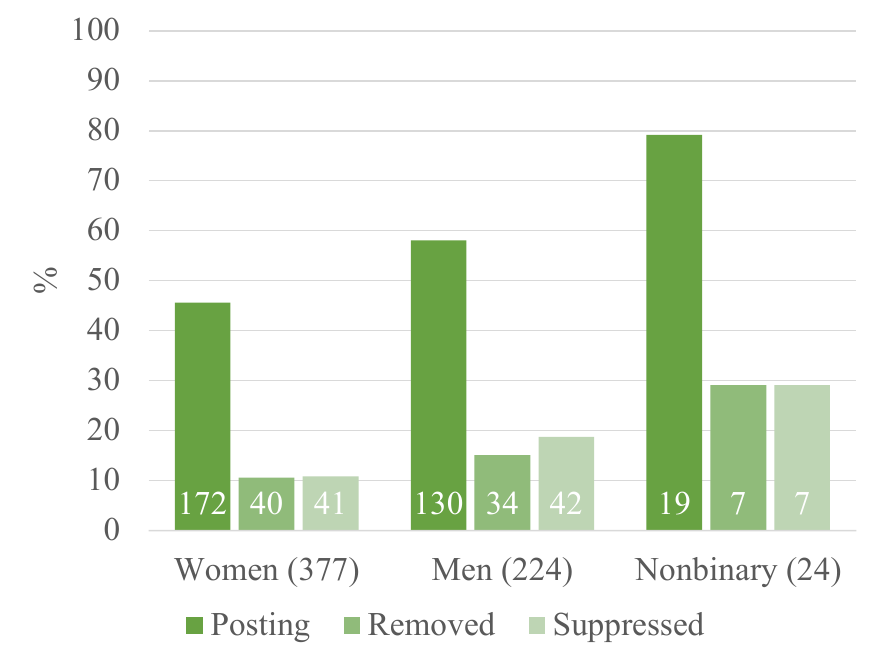}
\caption{Chart showing the percentage of respondents by gender who posted ``controversial'' content, and reported having content removed or suppressed. Data labels show counts. Total counts are given after identity labels. }
\label{fig:genderpercentage}
\end{figure}

Nonbinary respondents were the most likely to report posting one of the controversial topic types at least once, see Figure \ref{fig:genderpercentage}. 
Nonbinary people were more likely to post controversial content than men ($p=.050$) and women ($p=.001$). Nonbinary people were more likely to report having content removed and suppressed than men and women, see Figure \ref{fig:genderpercentage}
; the differences between nonbinary people and women are significant ($p=.014$ for removal, $p=.016$ for suppression). 

The types of content that respondents posted and reported as censored differed across genders. 
Men were significantly more likely to post political content than women, $28.6\%$, $n=64$, vs. $14.1\%$, $n=53$; $p<.001$, and more likely to have it removed ($11.76\%$ of those who reported content removal, $n=4$, compared to $n=0$ women, $p=.019$), though a Chi-square test with history of posting political content as a layer (control) variable suggests this was entirely accounted for by history of posting. 
Despite these differences, respondents of different genders shared similar beliefs about censorship of political content. All groups neither agreed nor disagreed if it was subject to algorithmic censorship ($3.15$ for women, $3.09$ for men, $3.25$ for nonbinary people), and all disagreed that it went against community guidelines ($2.45$ for women, $2.29$ for men, $2.15$ for nonbinary people). 
Gender did not impact ratings, per Kruskal-Wallis tests.

Men were much more likely to post content some may find offensive compared to women ($30.1\%$, $n=68$ for men; $11.7\%$, $n=44$ for women; $p<.001$), and more likely to report having this content removed ($7.59\%$, $n=17$ for men; $2.12$, $n=8$ for women; $p=.002$), 
though a Chi-square test 
suggests this was entirely accounted for by history of posting. Fifty percent of men who reported having content removed said it included content some may find offensive  ($n=17$). It was the most common type of content men reported having had removed. This was also the most commonly reported suppressed topic for men ($28.6\%$ of men who reported suppression, $n=12$). 
Similarly, men were significantly more likely to post hate speech than women. Over 5\% of men ($n=14$) reported posting hate speech at least once (compared with $1.9\%$ of women, $n=7$; $p=.016$). 


Nonbinary respondents were more likely than women to post content criticising a dominant group ($20.1\%$, $n=5$ vs. $6.37\%$, $n=24$; $p=.022$), related to a social justice movement ($30\%$, $n=8$ vs. $15.6\%$, $n=59$; $p=.042$) or about a minority experience ($58\%$, $n=5$ vs. $10.6\%$, $n=40$; $p<.001$). Nonbinary respondents were also significantly more likely to post about minority experiences than men 
(vs. $16.1\%$, $n=48$; $p<.001$). 


Turning to beliefs about censorship, we find that men, women and nonbinary people differ in their beliefs about the censorship of content about marginalised identities (criticising a dominant group, related to a social justice movement or about a minority experience). 
We find that nonbinary people agree more strongly on average that this content is algorithmically censored compared to women, who in turn believe more strongly in algorithmic censorship than men. We conducted Kruskal-Wallis tests (which are suitable for small sample sizes) and included pairwise comparisons with Bonferroni correction. We find respondents did not differ significantly in their beliefs about the algorithmic censorship of content criticising dominant groups, but did so for content related to social justice movements ($H(2)=19.909,p<.001$) and for content about minority experiences ($H(2)=21.347,p<.001$). Nonbinary people were significantly more likely than men to agree social justice content ($p=.001$) and minority experience content ($p=.001$) were algorithmically censored; women were also significantly more likely than men to agree social justice content ($p=.001$) and minority experience content ($p=.032$) were algorithmically censored; nonbinary people were significantly more likely than women to agree minority experience content was censored ($p=.002$). 

Our results suggest that there is not a clear relationship between history of censorship, and beliefs about algorithmic censorship. Nonbinary people did not differ significantly from men in terms of posting content related to a social justice movement, and there were no significant differences in reported removal or suppression for this topic (per Fisher's Exact Tests, $p=.209$ and $p=.186$ respectively), yet nonbinary people were more likely to agree this content is subject to algorithmic censorship. Likewise, beliefs about content suppression did not directly mirror content removal. For example, $14.3\%$ of men who reported having content suppressed reported this content was related to Covid ($n=6$), but less than $3\%$ reported having Covid-related content removed ($n=1$, $2.94\%$ of those who report content removal). 

Whilst respondents differed by gender in their beliefs about algorithmic censorship of content about marginalised identities, differences were not significant for beliefs about whether this content goes against community guidelines (per Kruskall-Wallis tests). Women, men and nonbinary people all disagreed that content related to a social justice movement or minority experience goes against community guidelines (for social justice content: $2.58$ for women, $2.60$ for men, $2.13$ for nonbinary people; for minority experience content: $2.68$ for women, $2.66$ for men, $2.22$ for nonbinary people). All three groups neither agreed nor disagreed if content criticising a dominant group goes against community guidelines ($3.02$ for women, $3.07$ for men, $2.89$ for nonbinary people). 





\subsubsection{By Trans Status}
Comparing trans and non-trans respondents, we found that trans respondents were much more likely to post controversial content ($75\%$ of trans respondents, $n=15$, vs. $50.6\%$, $n=303$; $p=.040$). However, trans respondents were no more likely to have content removed or suppressed ($p=.163$ and $p=.185$).

Looking at data on the types of content posted, one significant difference is in the frequency of posting about minority identity experiences which is (somewhat unsurprisingly) much higher for trans compared to non-trans respondents ($55\%$, $n=11$ vs. $12.7\%$, $n=76$; $p<.001$). Trans respondents were also comparatively more likely to report having this content removed ($20\%$ of those who report removal, $n=1$ vs. $2.7\%$, $n=6$; $p=.094$) or suppressed ($60\%$ of those who report suppression, $n=3$ vs. $7.2\%$, $n=6$; $p=.002$). Trans respondents also agreed more strongly that this content was subject to algorithmic censorship ($3.79$ vs. $3.31$; $p=.015$ per a Mann-Whitney U Test), although the two groups did not differ significantly in believing that this content does not go against community guidelines ($2.40$ for trans respondents and $2.67$ for non-trans respondents).


There were further differences in the reported beliefs about censorship and community guidelines: for example, trans respondents more strongly agreed that ``self-referential use of slur'' was censored by the algorithm ($3.96$ vs. $3.44$; Mann-Whitney U Test, $z=2.385$; $p=.017$). 




\subsubsection{By Sexuality}

\begin{figure}
\includegraphics[scale=0.45]{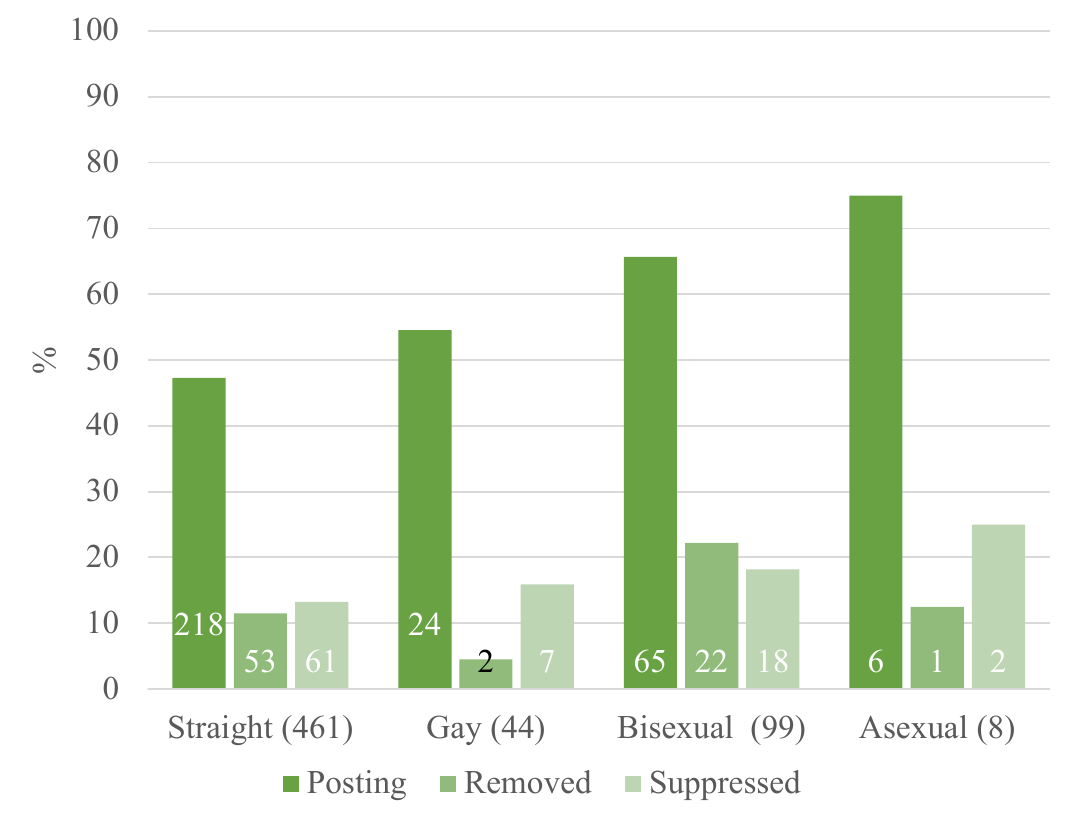}
\caption{Chart showing percentage of respondents by sexuality who posted ``controversial'' content, and reported having content removed or suppressed. Data labels show counts. Total counts are given after identities. }
\label{fig:sexualitypercentage}
\end{figure}

Counts for posting controversial content and reports of removal and suppression suggest bisexual and asexual respondents were the most likely to report these experiences, as shown in Figure \ref{fig:sexualitypercentage}. To establish whether likelihood of posting controversial content differed significantly across sexualities, we performed a cross-tabulation analysis with post-hoc $z$-tests across all sexualities to establish likely significant differences between sexualities. This suggested that bisexual respondents were more likely than straight respondents to post controversial content. We found this to be significant per a Fisher's Exact test  ($66.6\%$, $n=65$ vs. $47.3\%$, $n=218$; $p=.001$).

Repeating this procedure for content censorship, we find bisexual respondents were significantly more likely than straight respondents and gay respondents to report having had content removed. Fisher's Exact tests found these differences to be significant (compared to straight respondents: $22.2\%$, $n=22$ vs. $11.5$, $n=53$; $p=.012$; compared to gay respondents: vs. $4.54$, $n=2$; $p=.008$).

Whilst bisexual respondents greater posting of controversial content seems to explain the higher rates of censorship compared to straight respondents, this does not explain the comparative ``lack of censorship'' of gay respondents. Bisexual respondents report much higher rates of content removal compared to gay respondents, although a Chi-Square analysis with history of posting controversial content as a layer variable did not find this to be significant (for those with history of posting controversial content, $\chi^{2}(1,89)=3.771,p=.084$). Given the small sample sizes, we cautiously suggest this test may have been ``underpowered'' and that this result merits further investigation.



As with other identity groups, we did not find a clear link between reports of censorship and beliefs about algorithmic censorship. For example bisexual respondents were more likely to agree that ``sex related content for an erotic purpose'' was subject to algorithmic censorship, compared to straight respondents ($4.02$ for bisexual vs. $3.70$ for straight, per Mann-Whitney U test, $z=-2.349, p=.019$). However, there was no difference in posting rates between these two groups, nor did bisexual respondents report higher rates of censorship, suggesting this belief comes from observation of others' reports of censorship, or wider experience of the policing of queer sexualities, rather than direct experience of censorship on TikTok.

Straight respondents were most likely to report having content some may find offensive removed by the platform ($43.4\%$ of respondents who were straight and reported having content removed, $n=23$), but this was reported by only two bisexual respondents ($9.10\%$ of bisexual respondents who reported content being removed). A relatively large number ($n=6$) of straight respondents do \textit{not} report posting content that some may find offensive, but \textit{do} report having such content removed, perhaps indicative of a belief that the content they post is not offensive even if it is removed for being so. A comparison within those who report posting this type of content found bisexuals were less likely to have it removed than straight respondents, but this difference was not significant ($4.76\%$, $n=1$ vs. $23.0\%$, $n=17$; $p=.187$). 

LGB+ respondents were much more likely to post about minority identity experiences -- around $1/3$ of all gay, bisexual and asexual respondents report posting this type of content on average. Per a Fisher's Exact Test comparing straight and grouped non-straight (gay, bisexual, asexual) identities, this difference was significant $p<.001$. This was also the most common type of content for bisexual respondents who reported content suppression: $33.3\%$ ($n=6$) of bisexuals who reported experiencing suppression, equivalent to $1/5$ of all who reported posting this type of content. This contrasts with the fact only two bisexuals reported having this content removed ($9.10\%$ of bisexuals who reported content removal), reflecting a disconnect between experiences of content removal and beliefs about content suppression. 


\subsubsection{By Ethnicity}
Experiences of censorship differed by ethnic group  (See Table \ref{tab:ethniccontentremoval}). Black  respondents reported posting the most controversial content, and were significantly more likely to do so than White respondents ($66.7\%$, $n=30$ vs. $49.7\%$, $n=254$; $p=.030$), but no more likely than white respondents to report having content removed given a history of posting controversial content (per a Chi-Square analysis, $\chi^{2}(1,284)=0.120, p=.812$, for those who post). 

Almost a third of respondents of mixed or multiple ethnic groups reported content removal ($28.6\%$) and suppression ($33.3\%$); this is much higher compared to white respondents, even when controlling for history of posting controversial content: a Chi-square analysis found respondents from mixed or multiple ethnic groups were significantly more likely to report content suppression ($\chi^{2}(1,267)=4.399, p=.046$, for those who post).

All marginalised (non-white) ethnicities report higher rates of content suppression compared to removal ($\sim5\%$ greater or more) (see Table \ref{tab:ethniccontentremoval}). For Black respondents, reports of content suppression were almost double those of content removal, whereas for white respondents levels were similar. People of colour may perceive themselves to face higher levels of censorship through suppression, even though censorship through removal is typically reported at similar rates across ethnicities (barring mixed ethnicity). 

\begin{table}[]
\setlength\extrarowheight{2pt}
    \centering
    \begin{tabular}{wl{55pt}wl{45pt}wl{45pt}wl{45pt}}
    \hline
        Ethnic group & Posted & Removed & Suppressed \\ 
        \hline
         Asian & 48.9\% (22) & 13.3\% (6) & 20\% (9)  \\
        Black  & 66.7\% (30) & 11.1\% (5) & 20\% (9) \\ 
        Multiple & 61.9\% (13) & 28.6\% (6) &  33.3\% (7) \\
        White & 49.7\% (254) & 12.3\% (63) & 12.5\% (64) \\
    \hline
    \end{tabular}
    \caption{Percentage of respondents in each ethnic group who reported posting controversial content, having content removed or suppressed. Counts given in brackets.}
    \label{tab:ethniccontentremoval}
\end{table}




\subsubsection{By Disability}


\begin{figure}[t]
\includegraphics[scale=0.50]{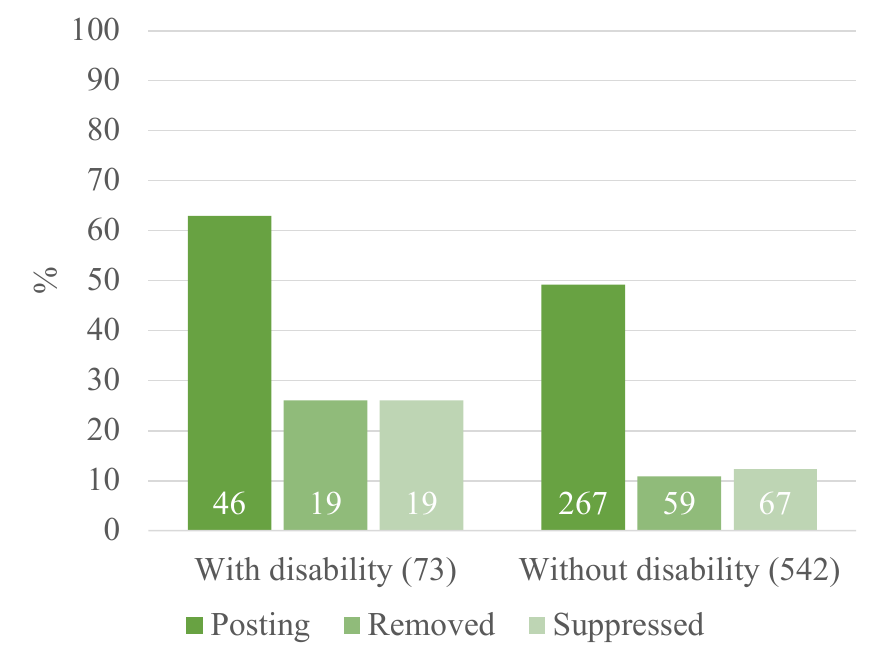}
\caption{Chart showing percentage of respondents by disability status who posted ``controversial'' content, and reported having content removed or suppressed. Data labels show counts. Total counts given after identities. }
\label{fig:disabilitypercentage}
\end{figure}

Respondents with disabilities were more likely to have content removed and suppressed, even when controlling for their greater likelihood to post controversial content (see Figure \ref{fig:disabilitypercentage}): Chi-square analyses with posting controversial content as a layer variable found respondents with disabilities to be significantly more likely to report content removal ($\chi^{2}(1,313)=13.259, p<.001$) and suppression ($\chi^{2}(1,313)=5.928, p=.022$).

Considering censorship of particular topics, as with LGBTQ+ minorities, respondents with disabilities were more likely to post about minority experiences compared to those without ($28.8\%$, $n=21$ vs. $12.2\%$, $n=66$; $p<.001$). Respondents with disabilities were more likely to report this content being removed ($n=3$ vs. $n=0$)
. A Chi-square analysis with posting about minority experiences as a layer variable found respondents with disabilities to be more likely to report removal of this content ($\chi^{2}(1,87)=9.765,p=.013$ for those who post). This may explain why respondents with disabilities seemed to more strongly agree that this content was subject to algorithmic censorship ($3.58$ vs. $3.30$), though this was not significant per a Mann-Whitney U Test ($z=1.795$; $p=.088$) -- although history of censorship did not always directly relate to types on content posted. For example, our results suggest that respondents with disabilities may agree more strongly that self-referential slurs are censored by the platform ($3.67$ vs. $3.43$) (though again this was not significant per a Mann-Whitney U Test, $z=1.752$; $p=.080$) though there was no difference in rates of censorship. 

\section{Discussion and Future Directions}
We identified a number of key trends in the data, which we discuss below. We also highlight avenues for future research to deepen our understanding of (perceived) biased censorship on TikTok. 

\paragraph{Offence is in the eye of the beholder}
We found that often content some may find offensive was one of the most commonly censored content types for non-marginalised groups (i.e. men and straight people). 
For example, straight people reported higher rates of removal than bisexual, gay or asexual people, and it was the most commonly removed content type; 
men reported higher rates of content removal than women and nonbinary people, and again it was the most commonly removed content type. Indeed, half of men who reported content being removed said it was content some may find offensive. This could relate to their greater tendency to post ``Telling jokes'' content compared to women and nonbinary people, which we also found, given the inherently subjective nature of humour. \citet{Meaney_Wilson_Chiruzzo_Magdy_2022} explore gender differences in rating online humour and find that men ``tolerate'' offence more in ``humorous'' content. It is also worth exploring the significantly higher rates of (self-reported) hate speech posting on the platform amongst men. 
It may be that men are no more likely to post hate speech, but are more likely to be honest about it. It may even be that men are less likely to take the topic seriously, so answered yes without much thought. Whatever the reason, it highlights the value of investigating hate speech content on TikTok, to understand who produces it and their attitudes towards offensive content. 

As with conservative TikTok users compared to trans and Black users in \citet{Haimson_Delmonaco_Nie_Wegner_2021}, it may be that whilst straight and male respondents report being censored at comparable rates to marginalised respondents, they are having content removed that (they themselves agree) is against community guidelines, where marginalised creators have content removed that does not clearly violate the platform's standards. Being censored for content that is believed to be in line with community guidelines will contribute towards feelings of discrimination and alienation. Users may feel they are being silenced for posting content that directly pertains to their lived experiences as marginalised people.  

\paragraph{Direct experience isn't everything}
We found beliefs about algorithmic censorship were not obviously correlated with direct experience of censorship on the platform. For example, nonbinary respondents agreed more strongly that social justice content is algorithmically censored, compared to men, even though nonbinary respondents never reported having this content removed. 
Thus, TikTok users' beliefs about censorship are not always the result of direct experience, but likely formed through which topics they (do not) see on the platform; the content they consume related to others' experiences of, and folk theories about, censorship; and censorship on other platforms and in ``real life''. 

\paragraph{Suspicious Minds}
We found there was a mismatch between reports of content removal, and content suppression. Around half of respondents who reported content suppression reported no content removal. Many more topics were reported as being subject to suppression compared to removal. Rates of suspected suppression were typically higher than rates of removal, across demographics. This suggests that beliefs about what content is suppressed by the platform are not directly related to experiences of content being removed (where a justification is typically provided). This is particularly evident when comparing across respondents from different ethnic groups. Reports of removal and suppression were at similar rates for white respondents, but respondents of colour were more likely to suspect suppression than removal. We do not attempt to say whether these suspicions are founded: it may be that users of colour are no more subject to suppression than white users, despite their beliefs. It may also be that TikTok favours suppression over removal of content by marginalised creators, to avoid similar scandals to \citet{Brown_2021, ghaffary_2021, ohlheiser_2021} \textit{i.a.} (whilst suppression of marginalised creators' content has also received significant media attention \citep{kelion_2019, kover_reuter_2019}, it is harder to prove than differing rates of content removal, account closures etc -- though cf. \citet{Ryan_Fritz_Impiombato_2020, Biddle_2020}. Experimentation in the style of \citet{King_Pan_Roberts_2014} -- who attempt to reverse engineer censorship in China -- may help to shed light on which topics are censored through suppression on the platform. However, regardless of the ``truth'', that some marginalised users feel they are subject to additional suppression is harmful in and of itself. 

\paragraph{Not So Innocent Mistakes}
We found respondents felt censorship of content did not always align with the platform's community guidelines. Respondents reported that content was removed without reason or that sometimes the ``wrong'' content was removed. We found that for several of the controversial topics (namely non-erotic sex related content, curse words, political content, content relating to minority identity experience and content relating to a social justice movement), respondents agreed this content was subject to algorithmic censorship despite not being against community guidelines. However, whilst respondents acknowledged that algorithmic censorship could go against official community guidelines, when asked about their beliefs about why content might be removed or suppressed despite not violating guidelines, respondents most strongly agreed that this was due to other users reporting the content. 
This aligns with previous research which found human intervention was considered a primary cause for content removal \citep{West_2018}. TikTok users have even expressed concerns this being done to harass creators \citep{Zeng_Kaye_2022, Are_2023}. 

\paragraph{The Bis Have It}
We found the bisexual respondents may be subject to additional censorship compared to gay respondents, though we repeat our caution that this requires further empirical investigation. This additional censorship of bisexual content may be best understood in the context of the finding of \citet{Simpson_Semaan_2021} that the TikTok algorithm seems to favour queer content that aligns with norms. Bisexual people face what is known as ``double discrimination'', whereby they face rejection from both outside and within the LGBTQ+ community for their plurisexual identity (attraction to multiple genders) \citep{Mereish_Katz-Wise_Woulfe_2017}. It is possible that content related to bisexual experiences falls outside the norm that has been constructed on TikTok -- by its developers and users -- hence bisexual respondents were subject to greater levels of censorship than gay respondents.

\section{Limitations}\label{sec:limitations}
We did not format our questions around TikTok's specific moderation guidelines (e.g. related to copyright music), which partly explains the high number of ``Other'' answers to types of censored content. This may have biased our findings against topics of particular relevance to TikTok such as minor safety. Our sampling method meant we obtained limited data from individuals with experience of censorship. An alternative method would be to target creators on TikTok who report censorship, or create videos to promote the study, but in both cases algorithmic curation would be a confound. 
Our use of the term ``censorship'' may have influenced responses, as censorship has more negative connotations than for example ``moderation''. Future work might employ more neutral language i.e. only using the terms ``removal'' and ``suppression''. We asked which types of content people post and which types are censored as separate questions, and some respondents gave conflicting answers. Mismatches may have arisen due to respondents preferring to select ``Never'' for the frequency of posting a type of content they have posted only once. Or respondents may have interpreted the content type descriptions differently in each instance, e.g. they do not believe the content they post to be sexual, but this was the reason given for its removal. Greater clarity in our instructions would have addressed this issue. 

Our sample may not be representative and we vehemently discourage use of the work to conclude that specific groups do not experience harms from censorship, simply because we did not find evidence. 


\section{Conclusion}
Our exploratory analysis has revealed that whilst rates of censorship are not always higher for marginalised creators, they are more likely to be censored for content that is generally regarded as in line with community guidelines, through removal and suppression. This will contribute towards feelings of discrimination and alienation on the platform. We have highlighted several avenues for future work into experiences of biased censorship on TikTok, to complement investigative work already conducted by journalists. We encourage a focus on users' beliefs over any attempt to identify the ``truth'', as these folk theories of censorship play a pivotal role in shaping users' experiences of the platform. 

\section{Acknowledgements}
Eddie L. Ungless is supported by the UKRI Centre for Doctoral Training in Natural Language Processing, funded by the UKRI (grant EP/S022481/1) and the University of Edinburgh, School of Informatics.

\bibliography{ref}
\clearpage

\includepdf[pages=-]{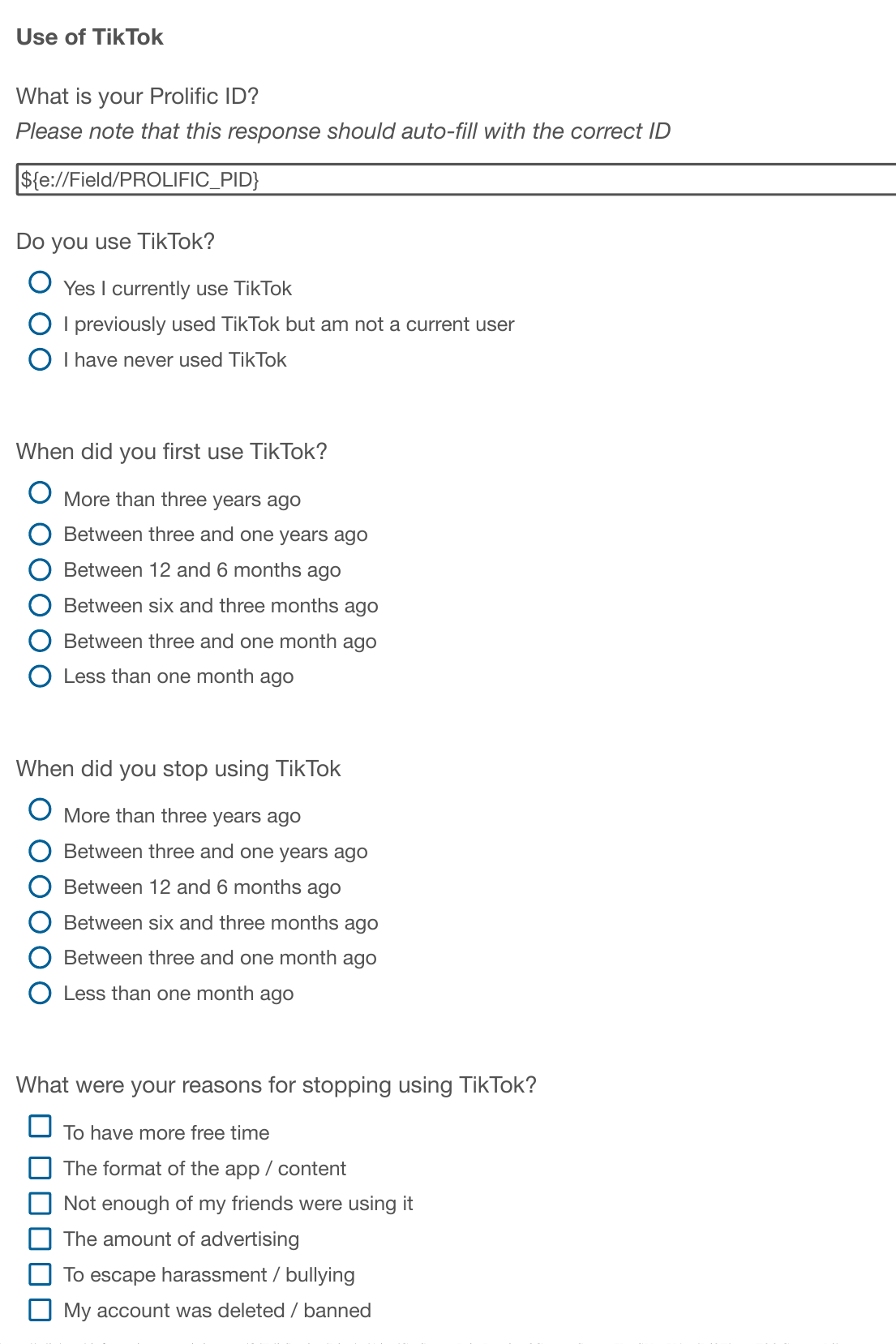}

\end{document}